\documentclass[reprint,superscriptaddress,showpacs,amsmath, amssymb,aps,pra]{revtex4-1}
\usepackage[T1]{fontenc}
\usepackage{amssymb}
\usepackage{graphicx}
\usepackage{dcolumn}
\usepackage{dsfont}
\usepackage{bm}
\usepackage{subfigure}
\usepackage[ruled]{algorithm2e}
\usepackage{hyperref}
\usepackage{appendix}

\hypersetup{colorlinks=true, citecolor=blue, urlcolor=blue, linkcolor=blue}
\begin{document}
\title{Quantum Information Masking in Non-Hermitian Systems and Robustness}
\author{Qiao-Qiao Lv}
\affiliation{School of Mathematical Sciences, Capital Normal University, 100048, Beijing, China}
\author{Jin-Min Liang}
\affiliation{School of Mathematical Sciences, Capital Normal University, 100048, Beijing, China}
\author{Zhi-Xi Wang}
\affiliation{School of Mathematical Sciences, Capital Normal University, 100048, Beijing, China}
\author{Shao-Ming Fei}
\affiliation{School of Mathematical Sciences, Capital Normal University, 100048, Beijing, China}

\begin{abstract}
\textbf{Abstract}

 By studying quantum information masking in non-Hermitian quantum systems, we show that mutually orthogonal quantum states can be deterministically masked, while an arbitrary set of quantum states cannot be masked in non-Hermitian quantum systems. We further demonstrate that a set of linearly independent states which are mutually $\eta$-orthogonal can be deterministically masked by a pseudo-unitary operator. Moreover, we study robustness of quantum information masking against noisy environments. The robustness of deterministic and probabilistic quantum information masking under different quantum noise channels is analyzed in detail. Accordingly, we propose and discuss the $r$-uniform probabilistic quantum information masking in multipartite systems.

\end{abstract}
\parskip=3pt

\maketitle

\section{Introduction}
Quantum correlations play important roles in quantum information processing \cite{hwang2003quantum,scarani2009the} and quantum computation \cite{Shor1994,Grover1996,HHL2009,Liang2020Variational,Liang2022eigenvalue}. A series of quantum no-go theorems such as quantum no-cloning \cite{wootters1982a}, quantum no-deletion \cite{pati2000impossibility} and quantum no-masking \cite{modi2018masking} are derived from the conventional of quantum mechanics \cite{schrodinger1926}.

The quantum information masking (QIM) introduced in the seminal work \cite{modi2018masking} has aroused widespread attention. It is closely related to quantum state discrimination \cite{ty2015}, information acceleration \cite{asn2021} and quantum secret sharing \cite{mva1999}. Besides deterministic versions, QIM also admits probabilistic realizations \cite{li2019deterministic,lms2020}. Recent works \cite{lms2018,hg2020,szc2021} have considered the QIM in multipartite systems and concluded that arbitrary quantum states can be masked when more participants take part in the masking process. For qubit case it has been shown that nonzero linear operators cannot mask a nonzero measure set of qubit states \cite{ll2019}. However, the qudit maskable set can have nonzero measure in $d$-dimensional Euclidean space \cite{lg2021}. Recently, quantum channel state masking \cite{uc2021}, hiding quantum information in real and  complex quantum system \cite{zhj2021} and  masking quantum information on hyperdisks \cite{dh2020} have been also investigated. From the aspect of resources, the randomness cost of masking quantum information can be thought of to be a meaningful method to quantify the quantum information masker \cite{lj2020}.

The no-go theorems above are based on the Hermitian quantum mechanics in which the evolution of a system is governed by the Schr\"{o}dinger's equation associated with an Hermitian Hamiltonian. The trace of density operator equals to one at arbitrary time and the expectation value of the energy with respect to a quantum state is real number. However, when a system interacts with environments, as an open quantum system it exhibits rich dynamical behaviors \cite{breuer2002the}. The interaction with the environment leads to a non-Hermitian Hamiltonian with a non-unitary evolution.

A natural question arises that if the no-go theorems in the conventional quantum mechanics still hold in non-Hermitian quantum systems. In \cite{ju2019non} the authors developed a modified formulation of non-Hermitian quantum mechanics (NHQM) based on the geometry of Hilbert spaces, in consistent with the conventional quantum mechanics for Hermitian systems. With the help of these formulations, it has been shown that the no-cloning and no-deleting theorems still hold in finite-dimensional non-Hermitian quantum systems. The quantum cloning and deleting in a pseudo-unitary system have been also studied \cite{chen2021quantum}.

In this work, based on the formulation of NHQM, we investigate the QIM in non-Hermitian Hamiltonian systems. We find that mutually orthogonal quantum states can be determinedly masked and an arbitrary quantum state cannot be masked in non-Hermitian quantum system. We further demonstrate that if a set of linearly independent states are mutually $\eta$-orthogonal, then the state set can be determinedly masked by a pseudo-unitary operator.
Moreover, since noises are inevitable in experimentally implementing QIM, quantum computing \cite{cehm1999} and quantum measurements \cite{cdjm2010}, we investigate the QIM under some families of Pauli channels. We present the definition of robustness of quantum information masking and demonstrate that the deterministic and probabilistic maskers are still efficient even at the presence of Pauli noise. Finally, we generalize the $r$-uniform QIM to probabilistic scenario.

\section{The NHQM Formalism}
We first briefly review the NHQM formalism introduced in \cite{ju2019non}. The dynamics of a state $|\Psi\rangle\rangle$ in non-Hermitian quantum mechanics is governed by the Schr\"{o}dinger equation,
\begin{equation}\label{1}
\begin{aligned}
\partial_t|\Psi\rangle\rangle=-\frac{i}{\hbar}\mathcal{H}|\Psi\rangle\rangle.
\end{aligned}
\end{equation}
Here the state $|\Psi\rangle\rangle$ and the conventional one $|\Psi\rangle$ in Hermitian quantum mechanics subject to a linear map,
\begin{equation}
\begin{aligned}
\langle\langle\Psi|=\langle\Psi|\eta,
\end{aligned}
\end{equation}
where $\eta$ is an Hermitian metric. The non-Hermitian Hamiltonian $\mathcal{H}$ satisfies the following relation,
\begin{equation}\label{e1}
\begin{aligned}
\mathcal{H}^{\dag}&=\eta \mathcal{H}\eta^{-1},
\end{aligned}
\end{equation}
where $\mathcal{H}^{\dag}$ is the transpose and conjugation of $\mathcal{H}$.
Similarly, a physical obserbvable $\mathcal{O}$ satisfies
\begin{equation}\label{e2}
\begin{aligned}
\mathcal{O}^{\dag}=\eta\mathcal{O}\eta^{-1}.
\end{aligned}
\end{equation}

Associated to a non-Hermitian system is the modified Hilbert space. The modified inner product of two states $|\Psi_1\rangle\rangle$ and $|\Psi_2\rangle\rangle$ is defined by,
\begin{equation}\label{e3}
\begin{aligned}
\langle\langle\Psi_1|\Psi_2\rangle\rangle_\eta=\langle\Psi_1|\eta|\Psi_2\rangle,
\end{aligned}
\end{equation}
with respect to an Hermitian operator $\eta$. When $\eta$ is  a positive-definite operator on the system Hilbert space $K$, i.e., $\langle x|\eta|x\rangle\geq0$ for all $|x\rangle\in K$ and it is 0 iff $|x\rangle=0$, it is easy to check that the formula (\ref{e3}) defines a definite positive inner product on $\mathcal{H}_S$ with which $\mathcal{H}_S$ becomes a Hilbert space, called a modified Hilbert space. We call such a $\eta$ a matric operator. Also, an operator $\mathcal{O}$ on $K$ is Hermitian (i.e., a physical observable) w.r.t. this new inner product (NIP) iff it satisfies Eq.(\ref{e2}). Especially, the Hamiltonian $\mathcal{H}$ is Hermitian w.r.t. the NIP due to the assumption (\ref{e1}). Moreover, an operator is said to be a generalized unitary operator (or a pseudo-unitary operator) if it preserves the NIP of any two vectors. Futhermore, a vector $|\psi\rangle\rangle$ in $K$ is a quantum state iff $\langle\psi|\eta|\psi\rangle=1$.

Due to the fact that the generalized unitary operator $\mathcal{U}$ does not change the $\eta$-inner product, one has
\begin{equation}
\begin{aligned}
\langle\langle\Psi_1|\Psi_2\rangle\rangle_\eta=\langle\langle\Psi_1|\mathcal{U}^{-1}\mathcal{U}|\Psi_2\rangle\rangle_\eta.
\end{aligned}
\end{equation}
Thus, the general unitary operator (pseudo-unitary operator) satisfies the following relation,
\begin{equation}
\begin{aligned}
\mathcal{U}^{\dag}=\eta\mathcal{U}^{-1}\eta^{-1}.
\end{aligned}
\end{equation}

\section{Quantum information masking in NHQM}
In what follows, we assume that $\mathcal{H}_A$ and $\mathcal{H}_B$ are two modified Hilbert spaces with metric operators $\eta_A$ and $\eta_B$, respectively. We say that an linear operator $\mathcal{S}$ form $\mathcal{H}_A$ into $\mathcal{H}_A\otimes \mathcal{H}_B$ makes a set of states $\{|\psi_k\rangle\rangle_{A}\}_{k=1}^m \subset \mathcal{H}_A$ as a set of states $\{|\Psi_k\rangle\rangle_{AB}\}_{k=1}^m \subset \mathcal{H}_A\otimes\mathcal{H}_B$ (i.e., $|\Psi_k\rangle\rangle_{AB}= \mathcal{S} |\psi_k\rangle\rangle_A, k=1,2, \ldots $). If there are two mixed states $\rho_A$ and $\rho_B$ of $\mathcal{H}_A$ and $\mathcal{H}_B$ such that
$$
\rho_A=\textrm{Tr}_B(|\Psi_k\rangle\rangle_{AB}\langle\langle\Psi_k|),~~\rho_B=\textrm{Tr}_A(|\Psi_k\rangle\rangle_{AB}\langle\langle\Psi_k|),
$$
for arbitrary $k$, we say such an $\mathcal{S}$ is a masker.


The masking is given by a pseudo-unitary operator $\mathcal{U}_{S}$ acting on the composite system $AB$ such that
\begin{equation}
\begin{aligned}
\mathcal{U}_{S}|a_k\rangle\rangle_A\otimes|b\rangle\rangle_B=|\Psi_{k}\rangle\rangle_{AB}.
\end{aligned}
\end{equation}
$\mathcal{U}_{S}$ gives a linear transformation and preserves the $\eta$-orthogonality. If two states $|\psi_1\rangle\rangle$ and $|\psi_2\rangle\rangle$ are $\eta$-orthogonal, then $\langle\langle\psi_1|\psi_2\rangle\rangle_{\eta}=0$.

\emph{Theorem 1.} No masker can mask all states of a modified Hilbert space $\mathcal{H}_A$ of dimension 2.

\emph{Proof.} Let us assume that $\mathcal{S}$ can mask all qubit states in $\mathcal{H}_A$. Let $\{|a_{0}\rangle\rangle_A,~|a_{1}\rangle\rangle_A\}$ be an $\eta$-orthogonal basis in the modified $\mathcal{H}_A$. The composite states $\{|\Psi_0\rangle\rangle_{AB},~|\Psi_1\rangle\rangle_{AB}\}$ are also $\eta$-orthogonal. An arbitrary qubit state $|a\rangle\rangle_A=\alpha_0|a_{0}\rangle\rangle_A+\alpha_1|a_{1}\rangle\rangle_A$ of system $A$ is mapped to the composite states $|\Psi\rangle\rangle_{AB}=\alpha_0|\Psi_{0}\rangle\rangle_{AB}
+\alpha_1|\Psi_{1}\rangle\rangle_{AB}$, $|\alpha_0|^2+|\alpha_1|^2=1$. The partial trace with respect
to the subsystems $A$ and $B$ are
\begin{equation}
\begin{aligned}
\textrm{Tr}_X(|\Psi\rangle\rangle_{AB}\langle\langle\Psi|)
&=|\alpha_0|^2\textrm{Tr}_X(|\Psi_0\rangle\rangle_{AB}\langle\langle\Psi_0|)\\
&+|\alpha_1|^2\textrm{Tr}_X(|\Psi_1\rangle\rangle_{AB}\langle\langle\Psi_1|)\\
&+\alpha_0^{*}\alpha_1\textrm{Tr}_X(|\Psi_1\rangle\rangle_{AB}\langle\langle\Psi_0|)\\
&+\alpha_0\alpha_1^{*}\textrm{Tr}_X(|\Psi_0\rangle\rangle_{AB}\langle\langle\Psi_1|),
\end{aligned}
\end{equation}
where $X=\{A,B\}$.
The masking conditions require that
\begin{equation}
\begin{aligned}
\alpha_0^{*}\alpha_1\textrm{Tr}_X(|\Psi_1\rangle\rangle_{AB}\langle\langle\Psi_0|)
+\alpha_0\alpha_1^{*}\textrm{Tr}_X(|\Psi_0\rangle\rangle_{AB}\langle\langle\Psi_1|)=0,
\end{aligned}
\end{equation}
for arbitrary $\alpha_0$ and $\alpha_1$. Then we have
\begin{equation}\label{ex1}
\begin{aligned}
\textrm{Tr}_X(|\Psi_1\rangle\rangle_{AB}\langle\langle\Psi_0|)=\textrm{Tr}_X(|\Psi_0\rangle\rangle_{AB}\langle\langle\Psi_1|)=0.
\end{aligned}
\end{equation}

Next we show that the above conditions cannot be satisfied for arbitrary qubit states. We use the following two $\eta$-orthogonal states to prove this result,
\begin{equation}\label{ex2}
\begin{aligned}
&|\Psi_0\rangle\rangle_{AB}=|\mu\rangle\rangle_A\otimes|a_{0}\rangle\rangle_B+|\nu\rangle\rangle_A\otimes|a_{1}\rangle\rangle_B,\\
&|\Psi_1\rangle\rangle_{AB}=|\mu_\bot\rangle\rangle_A\otimes|a_{0}\rangle\rangle_B+|\nu_\bot\rangle\rangle_A\otimes|a_{1}\rangle\rangle_B,
\end{aligned}
\end{equation}
where $|\mu\rangle\rangle$ and $|\nu\rangle\rangle$ are not necessarily mutually $\eta$-orthogonal and are not normalized, $|\mu\rangle\rangle$ and $|\mu_\bot\rangle\rangle$ are $\eta$-orthogonal, $|\nu\rangle\rangle$ and $|\nu_\bot\rangle\rangle$ are also $\eta$-orthogonal. Tracing over the subsystem $B$, we have the reduced states,
\begin{equation}\nonumber
\begin{aligned}
&\textrm{Tr}_B(|\Psi_0\rangle\rangle_{AB}\langle\langle\Psi_0|)=\beta_0 |\mu\rangle\rangle_{A}\langle\langle\mu|+
\beta_1 |\nu\rangle\rangle_{A}\langle\langle\nu|,\\
&\textrm{Tr}_B(|\Psi_1\rangle\rangle_{AB}\langle\langle\Psi_1|)=\tilde{\beta_0}|\mu_\bot\rangle\rangle_{A}\langle\langle\mu_\bot|+
\tilde{\beta_1}|\nu_\bot\rangle\rangle_{A}\langle\langle\nu_\bot|,\\
&\textrm{Tr}_B(|\Psi_0\rangle\rangle_{AB}\langle\langle\Psi_1|)=\tilde{\beta_0}|\mu\rangle\rangle_{A}\langle\langle\mu_\bot|+
\tilde{\beta_1}|\nu\rangle\rangle_{A}\langle\langle\nu_\bot|,\\
&\textrm{Tr}_B(|\Psi_1\rangle\rangle_{AB}\langle\langle\Psi_0|)=\beta_0|\mu_\bot\rangle\rangle_{A}\langle\langle\mu|+
\beta_1|\nu_\bot\rangle\rangle_{A}\langle\langle\nu|,
\end{aligned}
\end{equation}
where $\beta_0=\langle\langle a_{0}|a_{0}\rangle\rangle_{\eta_{B}}=\langle\langle a_{0}|a_{0}\rangle\rangle_{\eta_{B}}=\tilde{\beta_0}$, $\beta_1=\langle\langle a_{1}|a_{1}\rangle\rangle_{\eta_{B}}=\langle\langle a_{1}|a_{1}\rangle\rangle_{\eta_{B}}=\tilde{\beta_1}$, $\eta_B$ represents the Hermitian metric in space $\mathcal{H}_B$. From the masking conditions we have
\begin{equation}
\begin{aligned}
&\beta_0 |\mu\rangle\rangle_{A}\langle\langle\mu|+\beta_1|\nu\rangle\rangle_{A}\langle\langle\nu|\\
&=\tilde{\beta_0}|\mu_\bot\rangle\rangle_{A}\langle\langle\mu_\bot|+\tilde{\beta_1}|\nu_\bot\rangle\rangle_{A}\langle\langle\nu_\bot|.
\end{aligned}
\end{equation}
Computing the expectation value of the above equation with respect to $|\mu\rangle\rangle $, we obtain
\begin{equation}\label{ex3}
\begin{aligned}
\beta_0|\langle\langle\mu|\mu\rangle\rangle_{\eta_{A}}|^2+\beta_1|\langle\langle\mu|\nu\rangle\rangle_{\eta_{A}}|^2
=\tilde{\beta_0}|\langle\langle\mu|\nu_\bot\rangle\rangle_{\eta_{A}}|^2.
\end{aligned}
\end{equation}

Similarly, with respect to the expectation value of $\tilde{\beta_0}|\mu\rangle\rangle_{A}\langle\langle\mu_\bot|+
\tilde{\beta_1}|\nu\rangle\rangle_{A}\langle\langle\nu_\bot|$, according to (\ref{ex1}) we have
\begin{equation}
\begin{aligned}
\tilde{\beta_1}\langle\langle\mu|\nu\rangle\rangle_{\eta_{A}}
\langle\langle\nu_\bot|\mu\rangle\rangle_{\eta_{A}}=0.
\end{aligned}
\end{equation}
As $\tilde{\beta_1}\neq0$, we have either $\langle\langle\mu|\nu\rangle\rangle_{\eta_{A}}=0$ or $\langle\langle\nu_\bot|\mu\rangle\rangle_{\eta_{A}}=0$. Substituting the results to (\ref{ex3}), we obtain either
\begin{equation}\label{16}
\begin{aligned}
\beta_0|\langle\langle\mu|\mu\rangle\rangle_{\eta_{A}}|^2
=\tilde{\beta_0}|\langle\langle\mu|\nu_\bot\rangle\rangle_{\eta_{A}}|^2
\end{aligned}
\end{equation}
or
\begin{equation}\label{17}
\begin{aligned}
\beta_0|\langle\langle\mu|\mu\rangle\rangle_{\eta_{A}}|^2+\beta_1|
\langle\langle\mu|\nu\rangle\rangle_{\eta_{A}}|^2=0,
\end{aligned}
\end{equation}
where (\ref{17}) is obviously not true. Next we analyze (\ref{16}). According to $\langle\langle\mu|\mu\rangle\rangle_{\eta_{A}}\propto\langle\langle\mu|\nu_\bot\rangle\rangle_{\eta_{A}}$, we have $|\nu_\bot\rangle\rangle=e^{i\phi}|\mu\rangle\rangle$, from which and the inner product in $\textrm{Tr}_B(|\Psi_0\rangle\rangle_{AB}\langle\langle\Psi_1|)$ we obtain
\begin{equation}
\begin{aligned}
\tilde{\beta_0}\langle\langle\mu|\mu\rangle\rangle_{\eta_{A}}
\langle\langle\mu_\bot|\mu_\bot\rangle\rangle_{\eta_{A}}=0.
\end{aligned}
\end{equation}
Therefore, either $\langle\langle\mu|\mu\rangle\rangle_{\eta_{A}}=0$ or $\langle\langle\mu_\bot|\mu_\bot\rangle\rangle_{\eta_{A}}=0$. In this case  $|\Psi_0\rangle\rangle_{AB}$ and $|\Psi_1\rangle\rangle_{AB}$ are separable, which is a contradiction with (\ref{ex2}), we complete the proof. \hfill$\Box$

\emph{Theorem 2.} An arbitrary quantum state cannot be masked in non-Hermitian quantum system.

\emph{Proof.} Assume that a masker can mask two states $|s_0\rangle\rangle$ and $|s_1\rangle\rangle$, i.e.,
\begin{equation}
\begin{aligned}
&\mathcal{U}_{S}|s_0\rangle\rangle_A\otimes|b\rangle\rangle_B=|\Psi_0\rangle\rangle_{AB},\\
&\mathcal{U}_{S}|s_1\rangle\rangle_A\otimes|b\rangle\rangle_B=|\Psi_1\rangle\rangle_{AB},
\end{aligned}
\end{equation}
where $|\Psi_0\rangle\rangle_{AB},\,|\Psi_1\rangle\rangle_{AB}\in\mathcal{H}_A\otimes\mathcal{H}_B$. Then the same machine can mask the arbitrary superposed states $\mu|s_0\rangle\rangle_{A}+\nu|s_1\rangle\rangle_{A}$.
Since $|\Psi_0\rangle\rangle_{AB}$ and $|\Psi_1\rangle\rangle_{AB}$ are the purifications of $\rho_A^{(0)}$ and $\rho_A^{(1)}$, respectively, and $\rho_A^{(0)}=\rho_A^{(1)}$, they can be written as
\begin{equation}\label{sd}
\begin{aligned}
&|\Psi_0\rangle\rangle_{AB}=\sum_k\sqrt{\lambda_k}|a_k\rangle\rangle_A|b_k^{(0)}\rangle\rangle_B,\\
&|\Psi_1\rangle\rangle_{AB}=\sum_k\sqrt{\lambda_k}|a_k\rangle\rangle_A|b_k^{(1)}\rangle\rangle_B,
\end{aligned}
\end{equation}
where $\lambda_k$ are the eigenvalue of reduce density matrices, with the corresponding eigenvectors $\{|a_k\rangle\rangle_A\}_{k=1}^d$, $d=\min\{d_A,d_B\}$. By observation, $\{|a_k\rangle\rangle_A\}$ is an $\eta$-orthogonal set. Similarly, $\{|b_k^{(0)}\rangle\rangle_B\}$ and $\{|b_k^{(1)}\rangle\rangle_B\}$ are also $\eta$-orthogonal vectors, respectively.

From the definition of quantum information masking, we have
\begin{equation}
\begin{aligned}
\rho_B=\textrm{Tr}_A(|\Psi_0\rangle\rangle_{AB}\langle\langle\Psi_0|)=\textrm{Tr}_A(|\Psi_1\rangle\rangle_{AB}\langle\langle\Psi_1|).
\end{aligned}
\end{equation}
Assuming that the superposed state can be masked and taking the partial trace of the state $\mu|\Psi_0\rangle\rangle_{AB}+\nu|\Psi_1\rangle\rangle_{AB}$ with respect to $A$, we get
\begin{equation}
\begin{aligned}
\rho_B&=|\mu|^2\textrm{Tr}_A(|\Psi_0\rangle\rangle_{AB}\langle\langle\Psi_0|)+|\nu|^2\textrm{Tr}_A(|\Psi_1\rangle\rangle_{AB}\langle\langle\Psi_1|)\\
&\quad+\mu\nu^{*}\textrm{Tr}_A(|\Psi_0\rangle\rangle_{AB}\langle\langle\Psi_1|)\\
&\quad+\mu^{*}\nu\textrm{Tr}_A(|\Psi_1\rangle\rangle_{AB}\langle\langle\Psi_0|)\\
&=\rho_B+\mu\nu^{*}\textrm{Tr}_A(|\Psi_0\rangle\rangle_{AB}\langle\langle\Psi_1|)\\
&\quad+\mu^{*}\nu\textrm{Tr}_A(|\Psi_1\rangle\rangle_{AB}\langle\langle\Psi_0|).
\end{aligned}
\end{equation}
According to the masking conditions we have
\begin{equation}
\begin{aligned}
\mu\nu^{*}\textrm{Tr}_A(|\Psi_0\rangle\rangle_{AB}\langle\langle\Psi_1|)
+\mu^{*}\nu\textrm{Tr}_A(|\Psi_1\rangle\rangle_{AB}\langle\langle\Psi_0|)=0.
\end{aligned}
\end{equation}
Using (\ref{sd}) we have
\begin{equation}\label{ex4}
\begin{aligned}
\mu\nu^{*}\sum_k\lambda_k|b_k^{(0)}\rangle\rangle_B\langle\langle b_k^{(1)}|+\mu^{*}\nu\sum_k\lambda_k|b_k^{(1)}\rangle\rangle_B\langle\langle b_k^{(0)}|=0.
\end{aligned}
\end{equation}
Taking into account the expectation value of (\ref{ex4}) with respect to $|b_j^{(0)}\rangle\rangle$, we get
\begin{equation}
\begin{aligned}
\lambda_j(\mu\nu^{*}\langle\langle b_j^{(1)}|b_j^{(0)}\rangle\rangle_{\eta_{B}}+\mu^{*}\nu\langle\langle b_j^{(0)}|b_j^{(1)}\rangle\rangle_{\eta_{B}})=0.
\end{aligned}
\end{equation}
Since one can always choose $\lambda_j\geq 0$, the solutions are either $\mu=0$ or $\nu=0$ or $\langle\langle b_j^{(1)}|b_j^{(0)}\rangle\rangle_{\eta_B}=0$, where $\mu\nu^{*}\langle\langle b_j^{(1)}|b_j^{(0)}\rangle\rangle_{\eta_B}=0$ has purely imaginary solutions for all $j$, implying the restrictions on $\mu$ and $\nu$. \hfill$\Box$

\emph{Theorem 3.} For a set of mutually $\eta$-orthogonal sates $\{|a_{i}\rangle\rangle_{A}\}_{i=1}^{m}$, $|a_{i}\rangle\rangle_{A}\in\mathcal{H}_A$, $m\leq d=dim(\mathcal{H}_A)$, it can be deterministically masked by a pseudo-unitary operation, where $\eta$-orthogonal means that $\langle\langle a_{i}|a_{j}\rangle\rangle_{\eta}= g \delta_{ij}$ for all $i,j=1,2,\cdots,m$, where $ g \geq 0$.

\emph{Proof.} By observation a set of fixed reducing states $\{|\Psi_{k}\rangle\rangle_{AB}\}_{k\in \Gamma}$ can be written as
\begin{equation}
\begin{aligned}
|\Psi_{k}\rangle\rangle_{AB}=\Sigma_{i=1}^{d}\sqrt{\beta_{i}}|i\rangle\rangle_{A}|b_{i}^{(k)}\rangle\rangle_{B}.
\end{aligned}
\end{equation}
Especially, we consider the case where all the singular values are the same, i.e., $|\Psi_{k}\rangle\rangle_{AB}=\frac{1}{\sqrt{d}}\Sigma_{i=1}^{d}|i\rangle\rangle_{A}|b_{i}^{(k)}\rangle\rangle_{B}$.
Let
\begin{equation}
\begin{aligned}
\{|b_{i}^{(k)}\rangle\rangle_{B}\}_{i=1}^{m}=\{|\sigma_{m}^{k-1}(i)\rangle\rangle_{B}\}_{i=1}^{m},
\end{aligned}
\end{equation}
where $\sigma_{m}$ satisfying $\sigma_{m}(i)=i+1$ for $1\leq i \leq m-1$, and $\sigma_{m}(m)=1$. Since $\langle\langle i|j\rangle\rangle_{A}^{\eta}=g\delta_{ij}$, $i,j=1,2,\cdots,m$, one has $\langle\langle\Psi_{i}|\Psi_{j}\rangle\rangle_{AB}^\eta=g\delta_{ij}$. Based on the above analysis, we get a set of fixed reducing states $\{|\Psi_{k}\rangle\rangle_{AB}\}_{k=1}^{m}$ satisfying
\begin{equation}
\begin{aligned}
\langle\langle a_{i}|a_{j}\rangle\rangle_{A}^{\eta}\langle\langle b|b \rangle\rangle_{B}^{\eta}=\langle\langle a_{i}|a_{j}\rangle\rangle_{A}^{\eta}
=\langle\langle\Psi_{i}|\Psi_{j}\rangle\rangle_{AB}^{\eta}.
\end{aligned}
\end{equation}
for all $i,j=1,2,\cdots,m$.

From the fact that if two sets of states $\{|\phi_{i}\rangle\rangle\}_{i=1}^{m}$ and $\{|\varphi_{j}\rangle\rangle\}_{j=1}^{m}$ satisfy
\begin{equation}
\begin{aligned}
\langle\langle\varphi_{i}|\varphi_{j}\rangle\rangle_{\eta}=\langle\langle\phi_{i}|\phi_{j}\rangle\rangle_{\eta},
\end{aligned}
\end{equation}
then there exists a pseudo-unitary operator $\mathcal{U}$ such that $\mathcal{U}|\phi_{i}\rangle\rangle=|\varphi_{i}\rangle\rangle$ for $i=1,2,\cdots,m$,
so we get a pseudo-unitary operator $\mathcal{U}$ such that
\begin{equation}
\begin{aligned}
\mathcal{U}|a_{k}\rangle\rangle_{A}|b\rangle\rangle_{B}=|\Psi_{k}\rangle\rangle_{AB}.
\end{aligned}
\end{equation} \hfill$\Box$

According to the above theorems, we can crucially concluded that if $|\alpha_{1}\rangle\rangle$ and $|\alpha_{1}\rangle\rangle$ are not orthogonal in Hermitian quantum system, but are orthogonal in the sense of NHQM, then they can be deterministically masked in non-Hermitian quantum mechanics. In Ref. \cite{li2019deterministic} it has been proved that mutually orthogonal quantum states can be deterministically masked and linearly independent quantum states can be probabilistically masked. There are some states that are mutually $\eta$-orthogonal but not orthogonal in the Hermitian systems. We show that some states that are mutually $\eta$-orthogonal but not orthogonal in the Hermitian systems can be deterministically masked via a pseudo-unitary operator, see Fig. 1.
\begin{figure}[ht]
\includegraphics[scale=0.5]{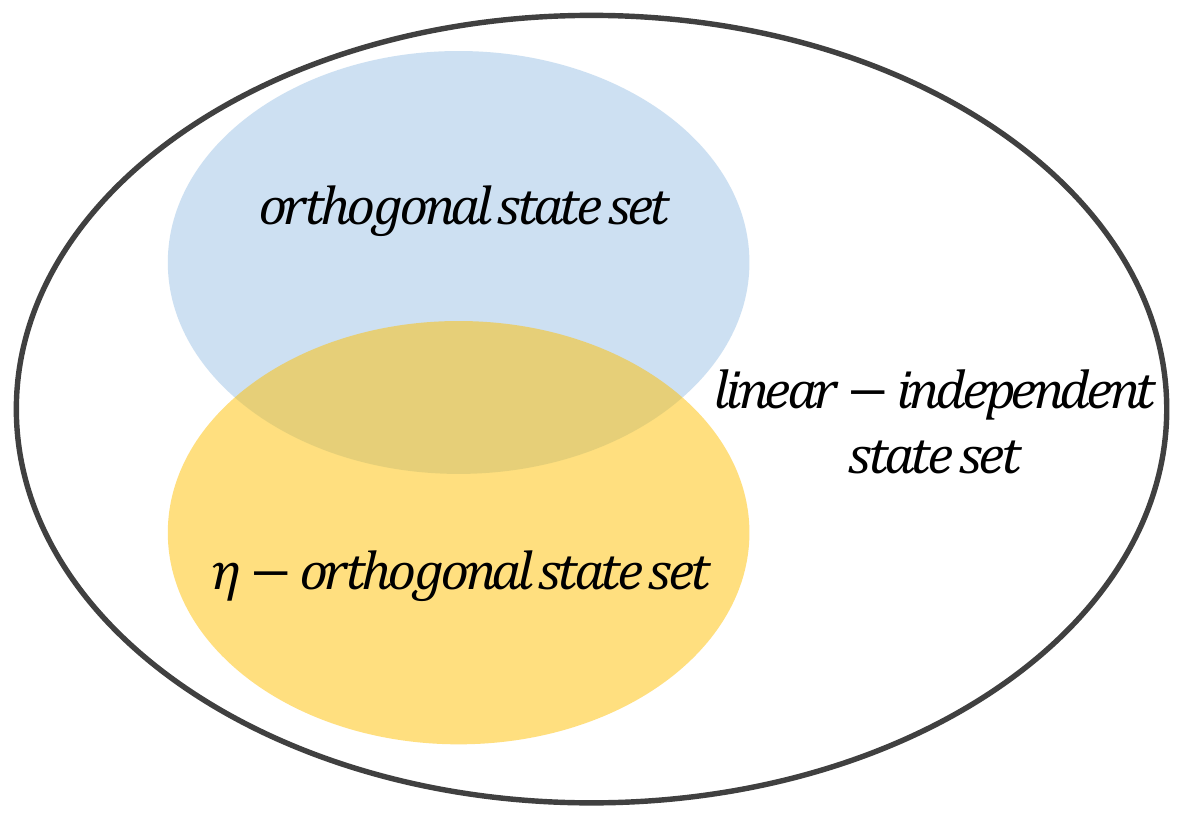}
\caption{The maskable sets of states: mutually orthogonal quantum states and linearly independent quantum states, as well as $\eta$-orthogonal states.}
\end{figure}

Let us two nonorthogonal quantum states $|\alpha_1\rangle$ and $|\alpha_2\rangle$ \cite{xp2020}:
\begin{equation}
\begin{aligned}
|\alpha_1\rangle=\frac{1}{\sqrt{\cosh\beta}}\left(\begin{array}{c}\cosh\frac{\beta}{2} \\[1mm] \sinh\frac{\beta}{2}\end{array}\right),\\
|\alpha_2\rangle=\frac{1}{\sqrt{\cosh\beta}}\left(\begin{array}{c}\sinh\frac{\beta}{2} \\[1mm] \cosh\frac{\beta}{2}\end{array}\right).
\end{aligned}
\end{equation}
One can easily verify that
\begin{equation}
\begin{aligned}
\langle\alpha_1|\alpha_2\rangle=\tanh\beta,~~
\langle\langle\alpha_1|\alpha_2\rangle\rangle_{\eta_0}=0,
\end{aligned}
\end{equation}
where $\eta_0 =a\mathds{I}-b\sigma_{x}$, $a= \cosh\beta$, $b=\sinh\beta$, $\sigma_{x}$ is Pauli matrix. Hence, $|\alpha_1\rangle$ and $|\alpha_2\rangle$ are non-orthogonal in conventional quantum mechanics, but are $\eta$-orthogonal in non-Hermitian system. According to \cite{li2019deterministic}, $|\alpha_1\rangle$ and $|\alpha_2\rangle$ can be probabilistically masked in Hermitian system, while based on our result they can be deterministically masked in non-Hermitian system.

\section{Robustness of quantum information masking}
In this section, we consider the quantum information masking under noise. In the quantum information masking one requires a unitary masker $U$ to prepare bipartite states such that the reduced states have no information about the maskable set of states. However, the noise in the process of applying the unitary $U$ may effect the information masking.

Let $\mathcal{E}$ be a quantum noise channel. Given a masker $\mathcal{S}$ with unitary $\mathcal{U}_{\mathcal{S}}$. We say that the (deterministic or probabilistic) quantum information masking is robust if and only if all the marginal states of $\mathcal{E}(\rho_{i}^{AB})$ are identical, where
$$\rho_{i}^{AB}=\mathcal{U}_{\mathcal{S}}\rho_{i}^{A}
\otimes\rho_{\textrm{anc}}^{B}\mathcal{U}_{\mathcal{S}}^{\dag},\quad i\in\{1,2,\cdots\},
$$
and $\rho_{\textrm{anc}}^{B}$ denotes an ancillary state.
We investigate in detail some quantum noise channels on deterministic and probabilistic masking.

\subsection{Robustness of deterministic quantum information masking}
For an arbitrary qudit state $\rho$, it can be expressed as
\begin{equation}
\begin{aligned}
\rho=\frac{1}{d}\mathds{I}_d+\frac{1}{2}\Sigma_{i=1}^{d^2-1}x_i\Lambda_i, \Sigma_{i=1}^{d^2-1}x_i^2\leq r_d^2, x_i\in \mathds{R},
\end{aligned}
\end{equation}
where $\Lambda_i$ is the base of $SU(d)$ group, $r_d$ satisfies $\frac{2}{d(d-1)}\leq r_{d}^{2} \leq \frac{2(d-1)}{d}$ \cite{K2003}.

As the case of $d=2$ and $r_d\equiv 1$, the states $\mathcal{D}_\alpha^\theta(\rho_0)$ in an arbitrary disk passing through the point $\rho_0=(x_0, y_0, z_0)$ in the Bloch sphere can be represented as
\begin{equation}
\begin{aligned}
\mathcal{D}_\alpha^\theta(\rho_0)=\{\rho: x\sin\alpha\cos\theta+y\sin\alpha\sin\theta+z\cos\alpha=c\},
\end{aligned}
\end{equation}
with $c=x_0\sin\alpha\cos\theta+y_0\sin\alpha\sin\theta+z_0\cos\alpha$, $\alpha\in[0,\pi]$ and $\theta\in [0,2\pi]$. They corresponding qubit masker $\mathcal{U}_\alpha^\theta$ has the form,
\begin{equation}\label{masker1}
\mathcal{U}_\alpha^\theta=
\left(
\begin{array}{cccc}
  \cos(\frac{\alpha}{2})&0& e^{-i\theta}\sin(\frac{\alpha}{2})&0\\
  0&\cos(\frac{\alpha}{2})&0&e^{-i\theta}\sin(\frac{\alpha}{2})\\
  0&\sin(\frac{\alpha}{2})&0&-e^{-i\theta}\cos(\frac{\alpha}{2})\\
  \sin(\frac{\alpha}{2})&0&-e^{-i\theta}\cos(\frac{\alpha}{2})&0
\end{array}
\right).
\end{equation}

After the Pauli channel, the single qubit state $\rho$ become the following form:
\begin{equation}
\begin{aligned}
\mathcal{E}_p^{P}(\rho)=p_{0}\rho+p_{1}X\rho X+p_{2}Y\rho Y+p_{3}Z\rho Z,
\end{aligned}
\end{equation}
where $X$, $Y$ and $Z$ are the standard Pauli matrices, $p_{0}+p_{1}+p_{2}+p_{3}=1$.
For $d$-dimensional Hilbert space, the Pauli channel can be represented by the \emph{Weyl channel} \cite{RB 2020},
\begin{equation}
\begin{aligned}
\mathcal{E}_p^{W}(\rho):=\Sigma_{s,t=0}^{d-1}p_{st}W_{st}\rho W_{st}^\dag,
\end{aligned}
\end{equation}
where $p_{st}$ are probabilities and the Weyl operators $W_{st}$ are given by
\begin{equation}
\begin{aligned}
W_{st}:=\Sigma_{m=0}^{d-1} e^{\frac{2\pi ims}{d}}|m\rangle\langle (m+t)\textrm{mod}(d)|.
\end{aligned}
\end{equation}

For $d=4$, we get, the specific form of the 4-dimension Pauli channel is given in the Appendix,
\begin{equation}
\begin{aligned}
\tilde{\rho}_{AB}&=\mathcal{E}_p^{W}(\mathcal{U}_\alpha^\theta\rho\otimes
|0\rangle\langle0|\mathcal{U}_\alpha^{\theta\dag})=\frac{1}{4}\mathds{I},
\end{aligned}
\end{equation}
where $\mathcal{U}_\alpha^\theta\rho\otimes|0\rangle\langle0|\mathcal{U}_\alpha^{\theta\dag}:=\rho_{AB}$
is given by
\begin{equation}\nonumber
\begin{aligned}
&\rho_{AB}=(1+x\sin\alpha\cos\theta+y\sin\alpha\sin\theta+z\cos\alpha)|0\rangle\langle0|\\
&+(-i x\sin\theta-x\cos\alpha\cos\theta-y\cos\alpha\sin\theta+iy\cos\theta)|0\rangle\langle3|\\
&+(ix\sin\theta-x\cos\alpha\cos\theta-y\cos\alpha\sin\theta-iy\cos\theta)|3\rangle\langle0|\\
&+(1-x\sin\alpha\cos\theta-y\sin\alpha\sin\theta-z\cos\alpha)|3\rangle\langle3|.
\end{aligned}
\end{equation}
If we select $p_{st}=\frac{1}{16}$, it is clear that $\textrm{Tr}_B(\tilde{\rho}_{AB})=\textrm{Tr}_A(\tilde{\rho}_{AB})$. In other words, the masker $\mathcal{U}_\alpha^\theta$ given by (\ref{masker1}) is still valid under Pauli noise.

As a concrete example, we consider the global depolarizing channel,$\mathcal{E}_q^{GD}$. The state of  $n$-qubit system after this channel is
\begin{equation}
\begin{aligned}
\mathcal{E}_q^{GD}(\rho)=(1-q)\rho+q\frac{\mathds{I}}{d},
\end{aligned}
\end{equation}
where $0\leq q \leq 1$, $d=2^n$.
Under the global depolarizing channel $\mathcal{E}_q^{GD}$, one has \cite{RB 2020},
\begin{equation}
\begin{aligned}
\tilde{\rho}_{AB}=\mathcal{E}_q^{GD}(\mathcal{U}_\alpha^\theta\rho\otimes|0\rangle\langle0|\mathcal{U}_\alpha^{\theta\dag}).
\end{aligned}
\end{equation}
By directly calculating, we get
$$\begin{aligned}
\rho_{A}=\textrm{Tr}_B(\tilde{\rho}_{AB})&=\frac{(1+c)(1-q)+q}{2}|0\rangle\langle0|\\
&+\frac{(1-c)(1-q)+q}{2}|1\rangle\langle1|,\\
\rho_{B}=\textrm{Tr}_A(\tilde{\rho}_{AB})&=\frac{(1+c)(1-q)+q}{2}|0\rangle\langle0|\\
&+\frac{(1-c)(1-q)+q}{2}|1\rangle\langle1|,
\end{aligned}$$
with $c=x_0\sin\alpha\cos\theta+y_0\sin\alpha\sin\theta+z_0\cos\alpha$, i.e. $\rho_{A}=\rho_{B}$. Therefore, the masker $\mathcal{U}_\alpha^\theta$ is robust again the the global depolarizing channel.

\subsection{Robustness of probabilistic quantum information masking}
A masker $\mathcal{S}_p$ probabilistically masks quantum information contained in states $\{|a_k\rangle_A\in\mathcal{H}_A\}$ to that in states $\{p_{k}|\Psi_k\rangle_{AB}\in\mathcal{H}_A\otimes\mathcal{H}_B\}$ such that all the marginal states of $|\Psi_k\rangle_{AB}$ are identical \cite{li2019deterministic,lms2020}. The probabilistically masker can be characterized by a completely positive and trace decreasing
linear map $L_{\mathcal{S}_{p}}$,
\begin{equation}
\begin{aligned}
\ L_{\mathcal{S}_{p}}|a_k\rangle_A\otimes|b\rangle_B=p_{k}|\Psi_{k}\rangle_{AB},
\end{aligned}
\end{equation}
where $|\Psi_{k}\rangle_{AB}$ satisfy
\begin{equation}
\begin{aligned}
\rho_A=\textrm{Tr}_B(|\Psi_k\rangle_{AB}\langle\Psi_k|), ~~ \rho_B=\textrm{Tr}_A(|\Psi_k\rangle_{AB}\langle\Psi_k|).
\end{aligned}
\end{equation}

The quantum information masking under Pauli and locally unitary noise channels has been considered for $n$-partite $d$-dimensional systems \cite{li2019deterministic}.
In \cite{h2021} the authors considered a special kind of quantum state, the $k$-uniform state, under particularly constructed Pauli channel. Here we extend the $r$-uniform QIM to probabilistic scenario and study general quantum states (pure or mixed) under more general quantum noise channels. We say that for multipartite systems a probabilistic operator $\mathcal{S}_p$ is an $r$-uniform quantum information probabilistic masker if
\begin{equation}
\begin{aligned}
\mathcal{S}_p : |a_k\rangle_{A_0} \rightarrow p_k|\Psi_k\rangle_{A_0,A_1,\cdots,A_{n-1}}
\end{aligned}
\end{equation}
such that $\textrm{Tr}_{A_{l_1},\cdots,A_{l_{n-r}}}(|\Psi_k\rangle_{A_0,A_1,\cdots,A_{n-1}}\langle \Psi_k|)$ are identical for $l_1,\cdots,l_{n-r} \in \{1,2,\cdots,n-1\}$.

The probabilistic masking process corresponds the linear transformation
$L_{\mathcal{S}_{p}}|k\rangle_{A_{0}}\otimes|b\rangle_{A_{1},\cdots,A_{n-1}}
=p_{k}|\Psi_{k}\rangle$, where $k=0,1,\cdots,d-1$, $0\leq p_k \leq 1$, $|b\rangle_{A_{1},\cdots,A_{n-1}}=|b_1\rangle\otimes|b_2\rangle\otimes\cdots|b_{n-1}\rangle$,  $\{|0\rangle,|1\rangle,\cdots,|d-1\rangle\}\in \mathds{C}^d$, $\{|\Psi_0\rangle,|\Psi_1\rangle,\cdots,|\Psi_{d-1}\rangle\}\in (\mathds{C}^d)^{\otimes n}$, and
\begin{equation}
\begin{aligned}
|\Psi_{k}\rangle=\frac{1}{\sqrt{d}}\Sigma_{j=0}^{d-1} e^{\frac{2\pi i}{d} jk}\underbrace{|j,j,\cdots,j\rangle}_{n}.
\end{aligned}
\end{equation}
One can easily check that all reduced matrices $(|\Psi_j\rangle\langle\Psi_j|)_{A_{l_1},\cdots,A_{l_m}}$ of $(|\Psi_j\rangle\langle\Psi_j|)_{A_{0},\cdots,A_{n-1}}$ are identical for $m\in \{1,2,\cdots,n-1\}$.

Under the $n$-partite $d$-dimensional Pauli channel $\mathcal{E}_p^{W}$ on $r$-uniform probabilistic masker $\mathcal{S}_p$, we get
\begin{equation}\label{pr}
\begin{aligned}
\mathcal{E}_p^{W}(p_k^2|\Psi_k\rangle\langle\Psi_k|)
=p_k^2 (q|\Psi_k\rangle\langle\Psi_k|+\frac{1-q}{d^n}\mathds{I}_{d^n}).
\end{aligned}
\end{equation}
The $r$-party reduced states of (\ref{pr}) are
\begin{equation}\label{R1}
\begin{aligned}
&\textrm{Tr}_{A_{l_1},\cdots,A_{l_{n-r}}}\Big[\mathcal{E}_p^{W}(p_k^2|\Psi_k\rangle\langle\Psi_k|)\Big]\\
&=p_k^2\Big(\frac{1}{d}\Sigma_{j=0}^{d-1}\underbrace{|j,j,\cdots,j\rangle}_{r}\underbrace{\langle j,j,\cdots,j|}_{r}+\mathds{I}_{d^{r}}\Big),
\end{aligned}
\end{equation}
where $A_{l_i} \in \{1,2,\cdots,n-r\}$. From (\ref{R1}) we see that probabilistic masking still works under Pauli noise. Moreover, the success probability is invariant. In other words, the state in $\{|0\rangle,|1\rangle,\cdots,|d-1\rangle\} \in \mathds{C}^d$ can still be probabilistically masked under Pauli noise.

\section{Conclusion and discussions}
In this work, we have investigated the quantum information masking in non-Hermitian quantum mechanics. We have demonstrated that a set of orthogonal states in the non-Hermitian Hamiltonian system can also be deterministically masked with a pseudo-unitary operator. Moreover, arbitrary set of qubit states can not be masked in the non-Hermitian Hamiltonian systems. However, we have shown that a probabilistic masker can mask a kind of linearly independent states (mutually $\eta$-orthogonal) in the conventional quantum mechanics deterministically by a pseudo-unitary operation in non-Hermitian Hamiltonian system.
Finally, we have proven that quantum deterministic and probabilistic masking still works under Pauli noise channels. In particular, the robustness of QIM enables people to hide quantum information under noisy environments, namely, although the correlation in the prepared states may be imperfect, one can still mask the maskable quantum states. We have only discussed the robustness of the conventional probabilistic quantum information masking under noise channels, the approaches can be also used to study the case of systems in non-Hermitian quantum mechanics.

\bigskip

Acknowledgements: This work is supported by the National Natural Science Foundation of China (NSFC) under Grant Nos. 12075159 and 12171044; Beijing Natural Science Foundation (Grant No. Z190005); Academy for Multidisciplinary Studies, Capital Normal University; Shenzhen Institute for Quantum Science and Engineering, Southern University of Science and Technology (No. SIQSE202001); Academician Innovation Platform of Hainan Province.

\section{APPENDIX}
The specific form of Weyl operators when $d=4$.
\begin{figure*}[]
\begin{small}
\begin{equation}
\begin{aligned}
&W_{00}=\left(\begin{array}{cccc}
  1&0&0&0\\
  0&1&0&0\\
  0&0&1&0\\
  0&0&0&1\end{array}\right),~~
W_{01}=\left(\begin{array}{cccc}
  0&1&0&0\\
  0&0&1&0\\
  0&0&0&1\\
  1&0&0&0\end{array}\right),~~
W_{02}=
\left(\begin{array}{cccc}
  0&0&1&0\\
  0&0&0&1\\
  1&0&0&0\\
  0&1&0&0\end{array}\right),\\
&W_{03}=\left(\begin{array}{cccc}
  0&0&0&1\\
  1&0&0&0\\
  0&1&0&0\\
  0&0&1&0\end{array}\right),~~
W_{10}=\left(\begin{array}{cccc}
  1&0&0&0\\
  0&i&0&0\\
  0&0&1&0\\
  0&0&0&i\end{array}\right),~~
W_{11}=\left(\begin{array}{cccc}
  0&1&0&0\\
  0&0&i&0\\
  0&0&0&-1\\
  -i&0&0&0\end{array}\right),\\
&W_{12}=\left(\begin{array}{cccc}
  0&0&1&0\\
  0&0&0&i\\
  -1&0&0&0\\
  0&-i&0&0\end{array}\right),~~
W_{13}=\left(\begin{array}{cccc}
  0&0&0&1\\
  i&0&0&0\\
  0&-1&0&0\\
  0&0&-i&0\end{array}\right),~~
W_{20}=\left(\begin{array}{cccc}
  1&0&0&0\\
  0&-1&0&0\\
  0&0&1&0\\
  0&0&0&-1\end{array}\right),\\
&W_{21}=\left(\begin{array}{cccc}
  0&1&0&0\\
  0&0&-1&0\\
  0&0&0&1\\
  -1&0&0&0\end{array}\right),~~
W_{22}=\left(\begin{array}{cccc}
  0&0&1&0\\
  0&0&0&-1\\
  1&0&0&0\\
  0&-1&0&0\end{array}\right),~~
W_{23}=\left(\begin{array}{cccc}
  0&0&0&1\\
  -1&0&0&0\\
  0&1&0&0\\
  0&0&-1&0\end{array}\right),\\
&W_{30}=\left(\begin{array}{cccc}
  1&0&0&0\\
  0&-i&0&0\\
  0&0&-1&0\\
  0&0&0&i\end{array}\right),~~
W_{31}=\left(\begin{array}{cccc}
  0&1&0&0\\
  0&0&-i&0\\
  0&0&0&-1\\
  i&0&0&0\end{array}\right),\\
&W_{32}=\left(\begin{array}{cccc}
  0&0&1&0\\
  0&0&0&-i\\
  -1&0&0&0\\
  0&i&0&0\end{array}\right),~~
W_{33}=\left(\begin{array}{cccc}
  0&0&0&1\\
  -i&0&0&0\\
  0&-1&0&0\\
  0&0&i&0\end{array}\right).
\end{aligned}
\end{equation}
\end{small}
\end{figure*}


\begin{thebibliography}{99}
\bibitem{hwang2003quantum} Hwang W Y 2003 Quantum Key Distribution with High Loss: Toward Global Secure Communication \emph{Phys. Rev. Lett.} \href{https://doi.org/10.1103/PhysRevLett.91.057901}{\textbf{91}, 057901}
\bibitem{scarani2009the} Scarani V, Pasquinucci H B, Cerf N J, Du\v{s}ek M, L\"{u}tkenhaus N and Peev M 2009 The security of practical quantum key distribution \emph{Rev. Mod. Phys.} \href{https://doi.org/10.1103/RevModPhys.81.1301}{\textbf{81}, 1301}
\bibitem{Shor1994} Shor P 1994 in \textit{Symposium on Foundations of Computer Science} (IEEE, Piscataway, NJ), pp. 124-134
\bibitem{Grover1996} Grover L K 1997 Quantum Mechanics Helps in Searching for a Needle in a Haystack \emph{Phys. Rev. Lett.} \href{https://doi.org/10.1103/PhysRevLett.79.325}{\textbf{79}, 325}
\bibitem{HHL2009} Harrow A W, Hassidim A and Lloyd S 2009 Quantum Algorithm for Linear Systems of Equations \emph{Phys. Rev. Lett.} \href{https://doi.org/10.1103/PhysRevLett.103.150502}{\textbf{103}, 150502}
\bibitem{Liang2020Variational} Liang J M, Shen S Q, Li M and Li L 2020 Variational quantum algorithms for dimensionality reduction and classification \emph{Phys. Rev. A} \href{https://doi.org/10.1103/PhysRevA.101.032323}{\textbf{101}, 032323}
\bibitem{Liang2022eigenvalue} Liang J M, Shen S Q, Li M and Fei S M 2022 Quantum algorithms for the generalized eigenvalue problem \emph{Quantum Inf. Process.} \href{https://doi.org/10.1007/s11128-021-03370-z}{\textbf{21}, 23}
\bibitem{wootters1982a} Wootters W K and Zurek W H 1982 A single quantum cannot be cloned \emph{Nature}  \href{https://doi.org/10.1038/299802a0}{\textbf{299}, 802}
\bibitem{pati2000impossibility} Pati A K and Braunstein S L 2000 Impossibility of deleting an unknown quantum state \emph{Nature} \href{https://doi.org/10.1038/404130b0}{\textbf{404}, 164}
\bibitem{modi2018masking} Modi K, Pati A K, Sen(De) A and Sen U 2018 Masking Quantum Information is Impossible \emph{Phys. Rev. Lett.} \href{https://doi.org/10.1103/PhysRevLett.120.230501}{\textbf{120}, 230501}
\bibitem{schrodinger1926} Schr\"{o}dinger E 1926 An Undulatory Theory of the Mechanics of Atoms and Molecules \emph{Phys. Rev.}  \href{https://doi.org/10.1103/PhysRev.28.1049}{28, 1049}
\bibitem{ty2015} Tian G J, Yu S X, Gao F, Wen Q Y and Oh C H 2015 Local discrimination of four or more maximally entangled states \emph{Phys. Rev. A} \href{https://doi.org/10.1103/PhysRevA.91.052314}{\textbf{91}, 052314}
\bibitem{asn2021} Abdelwahab A G, Ghwail S A, Metwally N, Mahran M H and Obada A S F 2021 The concealment of accelerated information is possible \emph{Quantum Inf. Process}. \href{https://doi.org/10.1007/s11128-021-03009-z}{\textbf{20}, 71}
\bibitem{mva1999} Hillery M, Bu\v{z}ek V and Berthiaume A 1999 Quantum secret sharing \emph{Phys. Rev. A}  \href{https://doi.org/10.1103/PhysRevA.59.1829}{\textbf{59}, 1829}
\bibitem{li2019deterministic} Li B, Jiang S H, Liang X B, Li-Jost X, Fan H and Fei S M 2019 Deterministic versus probabilistic quantum information masking \emph{Phys. Rev. A} \href{https://doi.org/10.1103/PhysRevA.99.052343}{\textbf{99}, 052343}
\bibitem{lms2020} Li M S and Modi K 2020 Probabilistic and approximate masking of quantum information \emph{Phys. Rev. A} \href{https://doi.org/10.1103/PhysRevA.102.022418}{\textbf{102}, 022418}
\bibitem{lms2018} Li M S and Wang Y L 2018 Masking quantum information in multipartite scenario \emph{Phys. Rev. A}  \href{https://doi.org/10.1103/PhysRevA.98.062306}{\textbf{98}, 062306}
\bibitem{hg2020} Han K Y, Guo Z H, Cao H X, Du Y X and Yang C 2020 Quantum multipartite maskers vs. quantum error-correcting codes \emph{Euro. Phys. Lett}
     \href{https://doi.org/10.1209/0295-5075/131/30005}{\textbf{131}, 30005}
\bibitem{szc2021} Shang W M, Zhang F L and Chen J L 2021 Quantum information masking basing on quantum teleportation \href{https://arxiv.org/abs/2103.03126}{arXiv:2103.03126}
\bibitem{ll2019} Liang X B, Li B and Fei S M 2019 Complete characterization of qubit masking \emph{Phys. Rev. A}  \href{https://doi.org/10.1103/PhysRevA.100.030304}{\textbf{100}, 030304(R)}
\bibitem{lg2021} Liu Z H, Liang X B, Sun K, Li Q, Meng Y, Yang M, Li B, Chen J L, Xu J S, Li C F and Guo G C 2021 Photonic Implementation of Quantum Information Masking \emph{Phys. Rev. Lett.}  \href{https://doi.org/10.1103/PhysRevLett.126.170505}{\textbf{126}, 170505}
\bibitem{uc2021} Pereg U, Deppe C and Boche H 2021 Quantum Channel State Masking \emph{IEEE Transactions on Information Theory}  \href{https://doi.org/10.1109/TIT.2021.3050529}{\textbf{10}, 1109}
\bibitem{zhj2021} Zhu H J 2021 Hiding and masking quantum information in complex and real quantum mechanics \emph{Phys. Rev. Research} \href{https://doi.org/10.1103/PhysRevResearch.3.033176}{\textbf{3}, 033176}
\bibitem{dh2020} Ding F and Hu X Y 2020 Masking quantum information on hyperdisks \emph{Phys. Rev. A} \href{https://doi.org/10.1103/PhysRevA.102.042404}{\textbf{102}, 042404}
\bibitem{lj2020} Lie S H and Jeong H 2020 Randomness cost of masking quantum information and the information conservation law \emph{Phys. Rev. A} \href{https://doi.org/10.1103/PhysRevA.101.052322}{\textbf{101}, 052322}
\bibitem{breuer2002the} Breuer H P and Petruccione F 2002 \href{https://doi.org/10.1093/acprof:oso/9780199213900.001.0001}{\emph{The theory of open quantum systems}} (Oxford University Press)
\bibitem{ju2019non} Ju C Y, Miranowicz A, Chen G Y and Nori F 2019 Non-Hermitian Hamiltonians and no-go theorems in quantum information \emph{Phys. Rev. A} \href{https://doi.org/10.1103/PhysRevA.100.062118}{\textbf{100}, 062118}
\bibitem{chen2021quantum} Chen Y C, Gong M, Xue P, Yuan H D and Zhang C J 2021 Quantum deleting and cloning in a pseudo-unitary system \emph{Front. Phys.} \href{https://doi.org/10.1007/s11467-021-1063-z}{\textbf{16}, 53601}
\bibitem{cehm1999} Cirac J I, Ekert A K, Huelga S F and Macchiavello C 1999 Distributed quantum computation over noisy channels \emph{Phys. Rev. A} \href{https://doi.org/10.1103/PhysRevA.59.4249}{\textbf{59}, 4249}
\bibitem{cdjm2010} Clerk A A, Devoret M H, Girvin S M, Marquardt F and Schoelkopf R J 2010 Introduction to quantum noise, measurement, and amplification \emph{Rev. Mod. Phys.} \href{https://doi.org/10.1103/RevModPhys.82.1155}{\textbf{82}, 1155}
\bibitem{xp2020} Zhan X, Wang K, Xiao L, Bian Z H, Zhang Y S, Sanders B C, Zhang C J and Xue P 2020 Experimental quantum cloning in a pseudo-unitary system \emph{Phys. Rev. A} \href{https://doi.org/10.1103/PhysRevA.101.010302}{\textbf{101}, 010302(R)}
\bibitem{K2003} Kimura G 2003 The Bloch vector for N-level systems \emph{Phys. Let. A} \href{https://doi.org/10.1016/S0375-9601(03)00941-1}{\textbf{10}, 1016}
\bibitem{RB 2020} LaRose R and Coyle B 2020 Robust data encodings for quantum classifiers \emph{Phys. Rev. A} \href{https://doi.org/10.1103/PhysRevA.102.032420}{\textbf{102}, 032420}
\bibitem{h2021} Hu M Y and Chen L 2021 Genuine entanglement, distillability and quantum information masking under noise \href{https://arxiv.org/abs/2102.00673}{arXiv:2102.00673}
\end{thebibliography}
\end{document}